\documentstyle[twoside,fleqn,espcrc2,psfig]{article}

\newcommand{\AmS}{{\protect\the\textfont2
  A\kern-.1667em\lower.5ex\hbox{M}\kern-.125emS}}
\newcommand{\be}{\begin{equation}}
\newcommand{\ee}{\end{equation}}
\newcommand{\ben}{\begin{eqnarray}}
\newcommand{\een}{\end{eqnarray}}

\def\simgt{\rlap{\lower 3.5 pt\hbox{$\mathchar \sim$}}\raise 1pt \hbox {$>$}}
\def\simlt{\rlap{\lower 3.5 pt\hbox{$\mathchar \sim$}}\raise 1pt \hbox {$<$}}

\title{$B_K$ with the Wilson Quark Action: A Non-Perturbative
       Resolution of Operator Mixing using Chiral
       Ward Identities\thanks{presented by Y.~Kuramashi}}

\author{JLQCD Collaboration \\ 
        S.~Aoki\address{Institute of Physics, University of
        Tsukuba, Tsukuba, Ibaraki 305, Japan},
        M.~Fukugita\address{Yukawa Institute for Theoretical 
        Physics, Kyoto University, Kyoto 606, Japan},
        S.~Hashimoto\address{National Laboratory for High
        Energy Physics (KEK), Tsukuba, Ibaraki 305, Japan},
        N.~Ishizuka$^{\rm a}$,
        Y.~Iwasaki$^{\rm a,}$\address{Institute of Physics, University of
        Tsukuba, Tsukuba, Ibaraki 305, Japan},
        K.~Kanaya$^{\rm a,d}$, Y.~Kuramashi$^{\rm c}$, 
        H.~Mino\address{Faculty of Engineering, Yamanashi
        University, Kofu 400, Japan},
        M.~Okawa$^{\rm c}$, A.~Ukawa$^{\rm a}$,
        T.~Yoshi\'{e}$^{\rm a,d}$}
       
\begin{document}

\begin{abstract}
We propose a non-perturbative method to determine the mixing
coefficients of $\Delta s=2$ four-quark operators for the
Wilson quark action using chiral Ward identities.
The method is applied to calculate $B_K$  in quenched QCD. 
\end{abstract}

\maketitle

\section{Introduction}

An essential step in the calculation of $B_K$ with the Wilson quark 
action is the resolution of the mixing problem of the $\Delta s=2$ 
four-quark operators, which is made difficult by the chiral symmetry 
breaking effects of the Wilson term.  
An apparent deficiency of perturbation theory for this problem has been 
well known\cite{lat88},  and most calculations have tried to 
resolve the mixing non-perturbatively with the aid of chiral perturbation 
theory\cite{bk_w}.  This method, however, has not been successful, 
since it contains large systematic uncertainties from higher order
effects which survive even in the continuum limit.
Recently the method of non-perturbative renormalization\cite{npr} has 
yielded a $\Delta s=2$ operator with a good chiral 
behavior\cite{romeBK}.  However, the underlying mechanism of improvement 
in this approach is not quite apparent.

Our aim is to calculate $B_K$ with a method which explicitly incorporates 
the chiral properties of the Wilson action, and to examine whether
the result is consistent with that using the Kogut-Susskind action.
In this report we propose a non-perturbative method to resolve the 
operator mixing problem based on chiral Ward identities(WI), and report 
first results of a calculation of $B_K$ carried out on VPP500/80 at KEK.

\section{Formulation of the method}

Let us consider a set of weak operators in the continuum  
$\{\hat O_i\}$ which closes under chiral rotation
$\delta^a \hat O_i=ic^a_{ij}\hat O_j$.  The continuum operators are
given by a linear combination of a set of lattice operators
$\{O_\alpha\}$, $\hat O_i=\sum_\alpha Z_{i\alpha}O_\alpha$.
We choose the mixing coefficients $Z_{i\alpha}$ such that
the Green's functions of $\{\hat O_i\}$ with quarks in
the external states satisfy the relevant chiral
Ward identities to $O(a)$.  The identities can be derived in a
standard manner\cite{wi} and take the form given by\\ \\[-2.5mm]
$-2\rho Z_A\langle\sum_xP^a(x)\hat O_i(0)
\prod_k\tilde\psi(p_k)\rangle$\\
$+c^a_{ij}\langle\hat O_j(0)\prod_k\tilde\psi(p_k)\rangle$\hfill (1)\\
$-i\sum_l\langle\hat O_i(0)\prod_{k\ne
l}\tilde\psi(p_k)\delta^a\tilde\psi(p_l)\rangle+O(a)=0$\\ \\[-2.5mm]
with $p_k$ the momentum of external quark.

The four-quark operator relevant for $B_K$ may be schematically written as
$\hat O_{VV+AA}=VV+AA$ with 
$V=\bar s\gamma_\mu d$ and $A=\bar s\gamma_\mu\gamma_5 d$.  Together
with $\hat O_{VA}=VA$, it forms a minimal set which closes
under $\lambda^3$ chiral rotation.   
The mixing pattern of these operators takes the form  
${\hat O_{VV+AA}}/2$=$Z_{VV+AA}\left(O_0+z_1O_1+ \cdots
+z_4O_4\right)$ and $O_{VA}=Z_{VV+AA}\cdot z_5O_5$ where the
lattice operators in the Fierz
eigenbasis are given by 
$O_0 = \left(VV + AA\right)/2$,
$O_1 = \left(SS+TT+PP\right)/2$,
$O_2 = \left(SS-TT/3+PP\right)/2$,
$O_3 = \left(VV-AA\right)/2 + \left(SS-PP\right)$,
$O_4 = \left(VV-AA\right)/2 - \left(SS-PP\right)$ 
and
$O_5=VA$.

Let us take four external quarks with an equal momentum
$p^2=\mu^2$.  Let $\Gamma_{VV+AA}$ and $\Gamma_{VA}$ be the sum of
Green's function on the left hand side of (1) with external
quark legs amputated.  Using the projection operator $P_i$ corresponding 
to $O_i$, we can write $
\Gamma_{VV+AA}=\Gamma_5P_5$ and $
\Gamma_{VA}=\Gamma_0P_0+\Gamma_1P_1+\cdots +\Gamma_4P_4$.  We then have 
six equations for the five coefficients $z_1,\cdots,z_5$,\\ \\[-2.5mm]
$
\Gamma_i/Z_{VV+AA}=
c^i_0+c^i_1z_1+\cdots +c^i_5z_5=O(a)\hfill (2)
$\\ \\[-2.5mm]
for $i=0,\cdots, 5$.
We may choose five equations to exactly vanish on the right
hand side. In the present analysis our choice is $i=1,\cdots,5$.
The remaining overall factor $Z_{VV+AA}$ is determined by the
non-perturbative renormalization method of ref.~\cite{npr}.
We convert final results for matrix elements into  those of the
$\overline{MS}$ scheme with naive dimensional
regularization (NDR) in the continuum at the renormalization scale
$\mu=2$GeV.

\section{Parameters of numerical simulation}

In Table 1 we summarize parameters of our simulations.
The lattice spacing is estimated from $m_\rho$.
At each $\beta$ we
employ four values of the  hopping parameter
such that the physical point for the $K$ meson can be
interpolated from data. We take degenerate $s$ and $d$ quark masses, 
and estimate $m_s a/2$ from $m_K/m_\rho=0.648$. 

For calculating Green's functions in Ward identities 
quark propagators are solved in the Landau gauge
for point source at the origin
with the periodic boundary condition. 
We extract
$B_K$ from a fit of plateau of the ratio of $K^0$-$\bar{K}^0$
Green's function of $\hat O_{VV+AA}$ divided by the vacuum
saturation of the $AA$ operator.  For this calculation
quark propagators are solved without gauge fixing for wall source at the 
edges of lattice using the Dirichlet boundary condition in the time
direction. Errors are estimated
by the single elimination jackknife procedure for all measured
quantities.

\begin{table}[t]
\vspace{-1mm}
\begin{center}
\caption{\label{tab:runpara}Run parameters.}
\begin{tabular}{llll}\hline
    $\beta$       & 5.9             & 6.1             & 6.3  \\ 
\hline
$L^3\times T$     & $24^3\times 64$ & $32^3\times 64$ & $40^3\times 96$  \\
\#conf.           & 300             & 100             & 50 \\
$a^{-1}$          & 1.95(5)      & 2.65(11)       & 3.41(20) \\ 
$m_s a/2$         & 0.0294(14)   & 0.0198(16)     & 0.0144(17) \\ 
\hline
\end{tabular} 
\end{center}
\vspace{-12mm}
\end{table}

\section{Results for $B_K$}

\begin{figure}[t]
\centering{
\hskip -0.0cm
\psfig{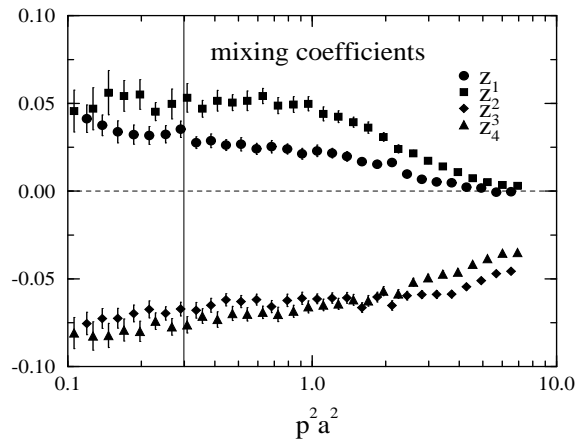}
\vskip -10mm  }
\caption{Mixing coefficients $z_1,\cdots,z_4$ 
for $K=0.15034$ at $\beta=6.3$. Vertical line corresponds to 
 $p\approx 2$GeV.}
\label{fig:bk_a_cl}
\vspace{-8mm}
\end{figure}

In Fig.~1 a representative result for the mixing coefficients is
plotted as a function of external quark momenta. Data show only a
weak scale dependence in the range  $0.2\simlt p^2a^2 \simlt 1.0$. We
take values of coefficients at $p\approx 2$GeV, which falls within
this range for our runs at $\beta=5.9-6.3$, in the following
analysis.  

We note that non-zero values for $z_2$ contrasts with the one-loop 
perturbative result $z_2=0$.  
Other coefficients agree in sign, albeit larger in magnitude, and 
approach perturbative values as $\beta$ increases.

\begin{figure}[t]
\centering{
\hskip -0.0cm
\psfig{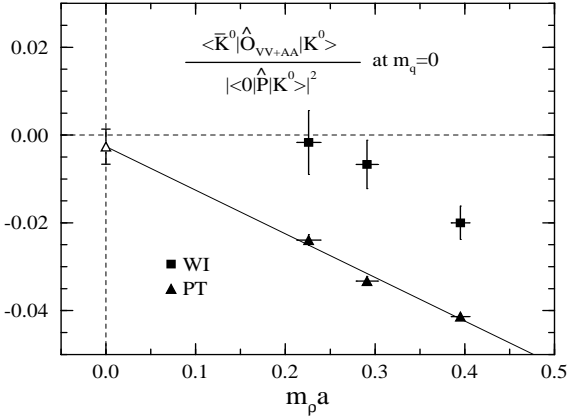}
\vskip -10mm  }
\caption{Effect of chiral symmetry breaking for Ward identity
(WI) and perturbative (PT) methods.}
\label{fig:bk_p_cl}
\vspace{-8mm}
\end{figure}
 
We show in Fig.~2 the $a$ dependence of the ratio $\langle
{\bar K}^0 \vert \hat O_{VV+AA} \vert K^0 \rangle /
\vert \langle 0 \vert {\hat P} \vert K^0 \rangle \vert^2$ 
extrapolated to $m_q=0$, which measures the contribution of chiral
symmetry breaking terms. Results are plotted both for our method (WI) and
with tadpole-improved one-loop perturbation theory (PT). The
scalar density $\hat P$ in the denominator is perturbatively
renormalized for both  cases.  A
significant improvement achieved with the use of Ward identities is
clearly seen, the ratio becoming consistent with zero even at
lattice spacing as large as $m_\rho a\approx 0.2-0.3$.

Within the one-loop resolution of operator mixing chiral breaking
effects are expected to appear as terms of $O(g^4)$  and
$O(a)$.  A roughly linear behavior of the PT values is consistent 
with the presence of the $O(a)$ term.  
Also they linearly  extrapolate to zero within errors at $a$$=$0.  
This suggests that 
$O(g^4)$ terms left out in the one-loop treatment are actually 
small.
 
Our results for $B_K({\rm NDR},2{\rm GeV})$ are summarized
in Fig.~3. The WI method gives reasonable values 
even at a finite lattice
spacing. Errors, however, are  large, and a continuum
extrapolation is difficult at this stage.  

In order to reduce statistical errors at each $\beta$, 
we employ an
alternative method(WI[VS]) in which the denominator of the
ratio for extracting $B_K$ is estimated from the vacuum
saturation of $\hat O_{VV+AA}$ constructed by the WI method.
While this method gives 
results different from those of WI at $a$$\ne$0, the
discrepancy is expected to vanish as $a$$\to$0.
A linear extrapolation in $a$ yields  $B_K({\rm
NDR},2{\rm GeV})=0.59(8)$.  This is consistent
with a recent JLQCD result for the Kogut-Susskind
action,
$B_K({\rm NDR},2{\rm GeV})=0.587(7)(17)$\cite{saoki}.

The tadpole-improved one-loop results (PT), if extrapolated linearly in $a$,
give $B_K$(NDR, 2GeV)=0.59(10), which agrees 
well those obtained with the WI or WI[VS] method. 
Uncertainties associated with a
large extrapolation has to be resolved for assessing the reliability 
of this approach, however.

In conclusion our results for $B_K$ demonstrates the effectiveness 
of the method of chiral Ward identities for constructing the $\Delta
s=2$ operator with the correct chiral property.  This makes us
hopeful to achieve the goal of a precision determination of
$B_K$  with the Wilson quark action with further improvement of our 
simulations.

\begin{figure}[t]
\centering{
\hskip -0.0cm
\psfig{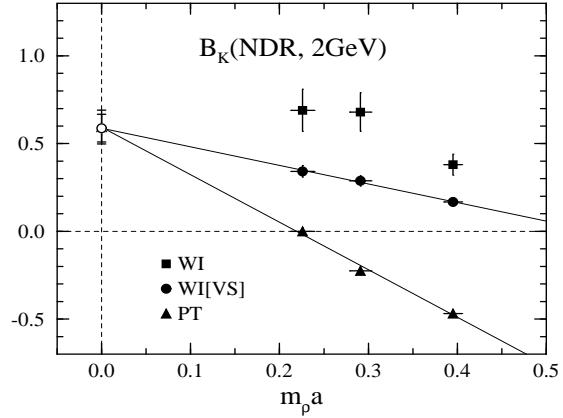}
\vskip -10mm  }
\caption{$B_K$ as a function of $a$.}
\vspace{-8mm}
\end{figure}

\end{document}